\documentstyle[12pt]{article}
\def\hybrid{\topmargin 0pt      \oddsidemargin 0pt
        \headheight 0pt \headsep 0pt
        \voffset=-0.5cm
        \textwidth 6.25in       % A4 paper
        \textheight 9.5in       % A4 paper
        \marginparwidth 0.0in
        \parskip 5pt plus 1pt   \jot = 1.5ex}
\catcode`\@=11
\def\marginnote#1{}

\newcount\hour
\newcount\minute
\newtoks\amorpm
\hour=\time\divide\hour by60
\minute=\time{\multiply\hour by60 \global\advance\minute by-\hour}
\edef\standardtime{{\ifnum\hour<12 \global\amorpm={am}%
        \else\global\amorpm={pm}\advance\hour by-12 \fi
        \ifnum\hour=0 \hour=12 \fi
        \number\hour:\ifnum\minute<10 0\fi\number\minute\the\amorpm}}
\edef\militarytime{\number\hour:\ifnum\minute<10 0\fi\number\minute}

\def\draftlabel#1{{\@bsphack\if@filesw {\let\thepage\relax
   \xdef\@gtempa{\write\@auxout{\string
      \newlabel{#1}{{\@currentlabel}{\thepage}}}}}\@gtempa
   \if@nobreak \ifvmode\nobreak\fi\fi\fi\@esphack}
        \gdef\@eqnlabel{#1}}
\def\@eqnlabel{}
\def\@vacuum{}
\def\draftmarginnote#1{\marginpar{\raggedright\scriptsize\tt#1}}

\def\draftlabel#1{{\@bsphack\if@filesw {\let\thepage\relax
   \xdef\@gtempa{\write\@auxout{\string
      \newlabel{#1}{{\@currentlabel}{\thepage}}}}}\@gtempa
   \if@nobreak \ifvmode\nobreak\fi\fi\fi\@esphack}
        \gdef\@eqnlabel{#1}}
\def\@eqnlabel{}
\def\@vacuum{}
\def\draftmarginnote#1{\marginpar{\raggedright\scriptsize\tt#1}}

\def\draft{\oddsidemargin -.5truein
        \def\@oddfoot{\sl preliminary draft \hfil
        \rm\thepage\hfil\sl\today\quad\militarytime}
        \let\@evenfoot\@oddfoot \overfullrule 3pt
        \let\label=\draftlabel
        \let\marginnote=\draftmarginnote
   \def\@eqnnum{(\theequation)\rlap{\kern\marginparsep\tt\@eqnlabel}%
\global\let\@eqnlabel\@vacuum}  }

\def\underline#1{\relax\ifmmode\@@underline#1\else
        $\@@underline{\hbox{#1}}$\relax\fi}

\def\titlepage{\@restonecolfalse\if@twocolumn\@restonecoltrue\onecolumn
     \else \newpage \fi \thispagestyle{empty}\c@page\z@
        \def\thefootnote{\fnsymbol{footnote}} }

\def\endtitlepage{\if@restonecol\twocolumn \else  \fi
        \def\thefootnote{\arabic{footnote}}
        \setcounter{footnote}{0}}  %\c@footnote\z@ }
\relax

\hybrid

\def\beq{\begin{equation}}
\def\eeq{\end{equation}}
\def\p{\partial}

\newtheorem{th}{Theorem}

\begin{document}

\begin{titlepage}

\title{Whitham-Toda hierarchy in the Laplacian
growth problem\footnote{Talk given at the Workshop NEEDS 99
(Crete, Greece, June 1999)}}

\author{M. Mineev-Weinstein \thanks{Theoretical Division, MS-B213, LANL,
  Los Alamos, NM 87545, USA }
\and A. Zabrodin
\thanks{Joint Institute of Chemical Physics, Kosygina str. 4, 117334,
Moscow, Russia and ITEP, 117259, Moscow, Russia}}

\date{October 1999}
\maketitle

\begin{abstract}

The Laplacian growth problem in the limit of zero
surface tension
is proved to be equivalent to
finding a particular solution to the dispersionless Toda
lattice hierarchy. The hierarchical times are harmonic
moments of the growing domain. The Laplacian growth equation
itself is the quasiclassical version of the
string equation that selects the solution to the hierarchy.

\end{abstract}

\vfill

\end{titlepage}

The Laplacian growth problem is one of the central
problems in the theory of pattern formation.
It has many different faces and a lot of important
applications.
In general words, this is about dynamics of
moving front (interface)
between two different phases. In many cases the dynamics
is governed by a scalar field that obeys the Laplace equation;
that is why this class of growth problems is called Laplacian.
Here we shall confine ourselves to the two-dimensional (2D)
case only.
To be definite, we shall speak about two
incompressible fluids with different viscosities on the plane.
In practice, the 2D geometry is
realized in the narrow gap between two plates.
In this version, this
is known as the Saffman-Taylor problem
or viscous fingering in the
Hele-Shaw cell.
For a review, see \cite{RMP}.

We shall mostly concentrate on the {\it external radial
problem} for it turns out to be the simplest case
in the frame of the suggested approach.
Let the exterior of a simply connected domain on the plane
be occupied by a viscous fluid (oil) while the interior
be occupied by a fluid with small viscosity (water).
The oil/water interface is assumed to be a simple
analytic curve. Other versions such as internal radial problem,
wedge or channel geometry are briefly discussed at the end
of the paper. Basically, they allow for the same approach.

Let $p(x,y)$ be the pressure, then $p$ is constant in the
water domain. We set it equal to zero.
In the case of zero surface tension $p$
is a continuous function across the interface, so
$p=0$ on the interface. In the oil domain
the gradient of $p$ is proportional to local velocity
$\vec V=(V_x,V_y)$ of the fluid
(Darcy's law):
\beq
\label{lg1}
\vec V=-\kappa \,\mbox{grad}\, p\,,
\eeq
where $\kappa$ is called the filtration coefficient\footnote{This
coefficient is inversely proportional to the viscosity, so the
Darcy law is formally valid in the water, too.}.
In particular, this law holds on the interface thus
governing its dynamics:
\beq
\label{lg1a}
V_n=-\kappa \frac{\p p}{\p n}\,.
\eeq
Here $V_n$ is the component of the velocity
normal to the interface
and $\p p/\p n$ is normal derivative.
Since the fluid is incompressible ($\mbox{div}\vec V =0$),
the Darcy law implies
that the potential $\Phi (x,y)=-\kappa p(x,y)$
is a harmonic function in the exterior (oil) domain:
\beq
\label{lg2}
\Delta \Phi (x,y)=0\,,
\eeq
where $\Delta =\p_{x}^{2}+\p_{y}^{2}$ is the Laplace operator
on the plane. The asymptotic behaviour of the function $\Phi$
very far away from the interface (at infinity) is determined
by the physical condition that there is a sink with constant
capacity $q$ placed at infinity. This means that
$$
\oint_{\gamma}(\vec V, \vec n)dl =q\,,
$$
where $\gamma$ is any closed contour encircling the water
domain,
$\vec n$ is the unit vector normal to $\gamma$.
So, we require $\Phi =0$ on the interface and
$\Phi=\frac{q}{2\pi}\mbox{log}\,|z|$ as $|z|\to \infty$.
The goal is to describe the interface motion subject
to the local dynamical law $V_n=\p \Phi /\p n$.

An effective tool for dealing with this
problem is the time dependent conformal
mapping technique (see e.g. \cite{RMP}).
Passing to the complex coordinates $z=x+iy$,
$\bar z =x-iy$ on the physical plane,
we bring into play a conformal map from
a reference domain on the mathematical plane $w$
to the growing domain on the physical plane.
By the Riemann mapping theorem, such a map does exist and,
under some conditions, is unique.
More precisely, let $z=\lambda (w)$ be the
univalent conformal map from
the exterior of the unit circle to the
exterior of the interface (i.e., to the oil domain)
such that $\infty$
is mapped to $\infty$ and the derivative $\lambda '(\infty)$
is a positive real number $r$. Under these conditions
the map is known to be unique.
The Laurent expansion of the $\lambda (w)$ around $\infty$
has then the following general form:
\beq
\label{z}
\lambda (w)=rw+\sum_{j=0}^{\infty}u_jw^{-j}\,.
\eeq

If the interface moves, the conformal map
$z(w,t)=\lambda (w,t)$ becomes time-dependent.
The interface itself is the image
of the unit circle $|w|=1$: as $w=e^{i\phi}$,
$0\leq \phi \leq 2\pi$, sweeps
over the unit circle, $z=\lambda (e^{i\phi},t)$ sweeps over the
interface at the moment $t$.

Having defined the conformal map, we immediately see that the
real part of the logarithm
of the inverse conformal map, $w(z)$,
provides the solution
to the Laplace equation in the oil domain with
the required asymptotics:
$\Phi(x,y)=\frac{q}{2\pi}\mbox{Re}\,\mbox{log}\,w(z)$.
Let us introduce
the complex velocity $V=V_x-iV_y$.
The obvious formula
$$
\frac{q}{2\pi}\frac{\p \,\mbox{log}\,w(z,t)}{\p z}=
\frac{\p \Phi}{\p x}-i\frac{\p \Phi}{\p y}
$$
allows one to represent
the Darcy law (\ref{lg1}) as follows:
\beq
\label{lg3}
V(z)=\frac{q}{2\pi}\,
\frac{\p \,\mbox{log}\,w}{\p z}\,,
\eeq
where the derivative is taken
at constant $t$.

In terms of the time-dependent conformal map,
the Darcy law is equivalent to the following
relation referred to as the Laplacian growth equation (LGE):
\beq
\label{lg6a}
\mbox{Im}\left (
\frac{\p z}{\p \phi}
\frac{\p \bar z}{\p t}\right )=\frac{q}{2\pi}\,.
\eeq
It first appeared in 1945 \cite{DAN} in the works on
the mathematical theory of oil production.
>From now on we set $q=\pi$ without loss of generality.
(This amounts to a proper rescaling of $t$.)
Introducing the Poisson bracket notation
\beq
\label{PB}
\{f,g\}=w\frac{\p f}{\p w}\frac{\p g}{\p t}
-w\frac{\p g}{\p w}\frac{\p f}{\p t}\,,
\eeq
for functions $f=f(w,t)$, $g=g(w,t)$ of $w$, $t$, we rewrite
the LGE in the suggestive form\footnote{Given a Laurent series
$f(z)=\sum_j f_j z^j$, we set
$\bar f(z)=\sum_j \bar f_j z^j$, so $z(w)$ and $\bar z(w^{-1})$
are complex conjugate only if $|w|=1$.}
\beq
\label{lg6}
\{z(w,t), \bar z(w^{-1},t)\}=1\,.
\eeq
The LGE thus means that the transformation from
$\mbox{log}\,w,\, t$ to $z, \bar z$ is canonical.

For a technical simplicity,
we assume that the point $z=0$ lies in the
water domain.
The interface dynamics given by the Darcy law (or,
equivalently, by the LGE) implies that if a point $(x,y)$ is
in the water (interior) domain at the initial moment,
then it remains there for all values of time. In particular,
our assumption that the point $z=0$ belongs to the water
domain means that zeros of the function $\lambda (w)$ are inside
the unit circle for any $t$.

The Laplacian growth is a particular case of
the 2D inverse potential problem. The shape of the
interface can be characterized by the
harmonic moments $C_k$ of the oil domain and the area
$C_0$ of the water domain:
\beq
\label{h1}
C_k=-\displaystyle{\int \!\!
\int_{\mbox{{\small exterior}}}}z^{-k}dxdy\,,
\;\;\;\;\;\; k\geq 1 ;\;\;\;\;\;C_{0}=\displaystyle{\int \!\!
\int_{\mbox{{\small interior}}}}dxdy\,.
\eeq
(The integrals at $k=1,2$ are assumed to be properly
regularized.) A remarkable result of Richardson \cite{Rich}
shows that
the LGE (\ref{lg6}) implies conservation of the harmonic
moments $C_k$ when the
interface moves, $dC_k/dt =0$, while area of the water
domain grows linearly in time: $C_0=\pi t$. Therefore, the problem
can be posed as follows: to find the shape of the domain
as a function of its area provided all the harmonic moments
of the exterior are
kept fixed. Since the harmonic moments
are coefficients in the expansion of the Coulomb
potential created by a homogeneously distributed charge
in the oil domain, this is just a
specification of the inverse potential problem.
We remind that to know the shape of the domain is the same as
to know the coefficients $r$, $u_j$ of the conformal map (\ref{z}).

A good deal of hints that the LGE has much to do
with integrable systems have
been known for quite a long time \cite{Rich}-\cite{M1}.
However, its status in the realm of integrability
was obscure until very recently.
The nature of the LGE and its relation to integrability
are clarified in the work \cite{MWZ}.
The idea is to treat the LGE {\it not
as a dynamical equation} but {\it as a constraint} in
a bigger integrable hierarchy.
The latter turns out to be an infinite
Whitham hierarchy of the type first
introduced in \cite{kri1}. This hierarchy is a
multi-dimensional generalization of integrable hierarchies of
hydrodinamic type \cite{hydro}. It naturally
incorporates the general inverse potential problem as well.
Namely, the coefficients of the conformal map as functions
of all the harmonic moments are given by a particular solution
to the dispersionless Toda
lattice hierarchy (see \cite{Tak-Tak} for a detailed
study of the latter). The hierarchical evolution
times are just the harmonic moments and their complex
conjugate. In the Laplacian growth all of them but the
area $C_0$ are frozen. Making them alive, one moves over
the space of initial data for the LGE, and recovers the
Whitham hierarchy.

For a more convenient formulation
of the result, let us rescale the harmonic
moments and introduce the new notation for them:
\beq
\label{11}
t\equiv t_{0}=\frac{C_0}{\pi},\;\;\;\;\;
t_k=\frac{C_k}{\pi k},\;\;\;\;\;\bar
t_k=\frac{\bar C_k}{\pi k},\;\;\;\;\; k\geq 1.
\eeq
The symbols $(f(w))_{\pm}$
below mean a truncated Laurent series, where
only  terms with positive (negative)
powers of $w$ are kept, $(f(w))_{0}$ is a
constant part ($w^0$) of the series.
\begin{th}
The conformal map (\ref{z}) obeys the following
differential equations with respect to the
harmonic moments:
\beq
\label{HJ9a}
\frac{\p \lambda (w)}{\p t_j} =\{H_j, \lambda (w)\}\,,
\;\;\;\;\;\;
\frac{\p \lambda (w)}{\p \bar t_j} =-\{\bar H_j, \lambda (w)\}\,,
\eeq
\beq
\label{HJ9b}
\frac{\p \bar \lambda (w^{-1})}{\p t_j}
=\{H_j, \bar \lambda  (w^{-1})\}\,,
\;\;\;\;\;\;
\frac{\p \bar \lambda (w^{-1})}{\p \bar t_j} =
-\{\bar H_j, \bar \lambda  (w^{-1})\}\,,
\eeq
where the Poisson bracket is defined in (\ref{PB}), and
\beq
\label{HJ12}
H_j(w)=\Bigl (\lambda ^j(w)\Bigr )_{+} +\frac{1}{2}
\Bigl (\lambda ^j(w) \Bigr )_{0}\,,
\eeq
\beq
\label{J4}
\bar H_j(w)=\Bigl (\bar \lambda ^j(w^{-1})\Bigr )_{-} +\frac{1}{2}
\Bigl (\bar \lambda ^j(w^{-1}) \Bigr )_{0}\,.
\eeq
\end{th}
The proof is sketched in
\cite{MWZ} and \cite{WZ}. These are the Lax-Sato
equations for the dispersionless Toda lattice hierarchy of
non-linear differential equations, with the $\lambda (w)$
and $\bar \lambda (w^{-1})$ being the Lax functions.
On comparing coefficients
in front of powers of $w$ in (\ref{HJ9a}), (\ref{HJ9b}),
one obtains an infinite set of non-linear differential
equations for the coefficients of the comformal map. Altogether,
they form the hierarchy. The particular solution that solves the
inverse potential problem is selected by the constraint
(\ref{lg6}), where $z(w,t)=\lambda (w,t)$,
$\bar z(w,t)=\bar \lambda (w^{-1},t)$,
and all $t_k$ are fixed.
This constraint is known as (a quasiclassical version of) the
{\it string equation}. Surprisingly, this very constraint
is the key ingredient of the integrable structures in
2D gravity coupled with $c=1$ matter \cite{gravity,matrix}.
The mathematical theory of the dispersionless hierarchies
constrained by string equations was developed in
\cite{kri2,Du1} and
extended to the Toda case in \cite{Tak-Tak}.

The Lax-Sato equations imply \cite{kri2}
the existence of the prepotential function
$F(t, t_k, \bar t_k)$. This function
solves the inverse potential problem
in the following sense.
Let $C_{-k}$, $k\geq 1$, be the complimentary set of
harmonic moments,
i.e., the moments of the interior of the domain:
\beq
\label{h2}
C_{-k}=\displaystyle{\int \!\!
\int_{\mbox{{\small interior}}}}z^{k}dxdy\,.
\eeq
Were we able to find $C_{-k}$ from a given set $C_0$,
$C_k$ (i.e., given $t$, $t_k$), this would yield the complete
solution, alternative but equivalent
to knowing the conformal map (\ref{z}). Indeed, using
the contour integral representation of the harmonic
moments, it is easy to see that
the generating function of {\it all} the harmonic moments,
obtained as an analytic continuation of the Laurent series
\beq
\label{S}
\mu (\lambda )= \frac{1}{\pi}
\sum_{k\in {\bf Z}}C_k \lambda ^k =
\sum_{k=1}^{\infty}kt_k \lambda ^{k}+t+
\frac{1}{\pi}\sum_{k=1}^{\infty}C_{-k} \lambda ^{-k}\,,
\eeq
would allow us to
restore the interface curve via the equation
$|z|^2 =\mu (z)$ ($S(z)= \mu (z)/z$ is what is called
{\it Schwarz function} of the curve \cite{Davis}).
The function $F$ does the job.
\begin{th}
(\cite{MWZ,WZ})
There exists a real function $F$ of the (rescaled) harmonic
moments $t$, $t_k$, $\bar t_k$ such that
\beq
\label{F}
C_{-k}=\pi \,\frac{\p F}{\p t_k}\,,
\;\;\;\;\;\;
\bar C_{-k}=\pi \,\frac{\p F}{\p \bar t_k}\,,
\;\;\;\;\; k\geq 1\,.
\eeq
\end{th}
In particular, this implies the symmetry of
derivatives of the inner harmonic
moments with respect to the outer ones:
\beq
\label{sym}
\frac{\p C_{-k}}{\p t_j}=
\frac{\p C_{-j}}{\p t_k}\,,
\;\;\;\;\;\;
\frac{\p C_{-k}}{\p \bar t_j}=
\frac{\p \bar C_{-j}}{\p t_k}\,.
\eeq
For general Whitham hierarchies, the function $F$
was introduced in \cite{kri2}.
In the dispersionless Toda case, it is the
dispersionless limit of logarithm of the
$\tau$-function \cite{Tak-Tak}.
This function obeys the dispersionless limit of the Hirota
equation (a leading term of the differential Fay identity
\cite{GiKo1},\,\cite{Tak-Tak}):
\begin{equation}
\label{Hirota}
(z\! -\! \zeta)\,\exp \! \left(\sum_{n,m\geq
1}\frac{F_{nm}}{nm}z^{-n}\zeta^{-m}\right)\! =
\! z\exp \! \left(-\sum_{k\geq
1}\frac{F_{0k}}{k}z^{-k}\right)\! -\!\zeta\exp \! \left(-\sum_{k\geq
1}\frac{F_{0k}}{k}\zeta^{-k}\right)
\end{equation}
This is an infinite set  of relations between the second derivatives
$F_{nm}=
\p_{t_n}\p_{t_m}F,\;\;\;F_{0m}=
\p_{t}\p_{t_m}F$.
The equations   appear  while expanding the
both sides  of (\ref{Hirota}) in
powers of $z$ and $\zeta$.

To complete the identification with the objects of the
theory of Whitham hierarchies, we just mention that
the function $\mu (\lambda )$ (\ref{S}) is the quasiclasical limit
of the Orlov-Shulman operator \cite{Or} for the 2D Toda lattice.
This function obeys the conditions
$\{\lambda , \mu \}=\lambda$,
$\{\bar \lambda , \bar \mu \}=-\bar \lambda$.
The string equation can be then written as a relation
\beq
\label{string1}
\bar \lambda =\lambda ^{-1} \mu
\eeq
between the Laurent series, together with the
reality condition $\mu (\lambda )=\bar \mu (\bar \lambda)$.
Written in this form,
it admits generalizations which are also relevant
to the Laplacian growth
problem, and which we are going to discuss now.

Consider the Laplacian growth problem for domains
symmetric under the ${\bf Z}_n$-\-trans\-for\-ma\-ti\-ons
$z\to e^{2\pi i l/n}z$, $l=1,\ldots , n-1$, where
$n$ is a positive integer. In this case $C_m=0$
unless $m=0 \,(\mbox{mod}\,n)$, i.e., only the moments
$C_{kn}$ are non-zero.
The conformal map from the exterior of the unit circle
to the exterior of such a domain
is given by
$z(w)=(\lambda (w^n))^{1/n}$,
where $\lambda (w)$ has the same general structure as
in (\ref{z}).
Equivalently, the problem may be
posed as the Laplacian growth in the wedge (sector) domain
restricted by the rays $\mbox{arg}\,z =0$ and
$\mbox{arg}\,z =2\pi /n$ with the periodic condition
on the boundary. In the latter case, the conformal map is
\beq
\label{conf2}
z(w)=(\lambda (w))^{1/n}\,.
\eeq
The LGE in terms of the conformal map
$z(w,t)$ has the same form (\ref{lg6}) if the time
is rescaled as $t \to t/n$.
In terms of the $\lambda (w,t)$, one has
\beq
\label{lg61}
\{\lambda ,\, \bar \lambda \}=
n^2(\lambda \bar \lambda )^{\frac{n-1}{n}}\,.
\eeq

Let us introduce the following notation for the
non-zero harmonic moments:
\beq
\label{111}
t=\frac{C_{0}}{n\pi },\;\;\;\;\;
t_k=\frac{C_{kn}}{\pi k},\;\;\;\;\;
\bar t_k=\frac{\bar C_{kn}}{\pi k},\;\;\;\;\; k\geq 1
\eeq
(cf.\,(\ref{11})).
The more general string equation (\ref{lg61}) amounts to
the relation
\beq
\label{string2}
\bar \lambda =n^n \lambda ^{-1} \mu ^n
\eeq
between the Lax and Orlov-Shulman functions (note that
this means $|z|^2 =n\mu (z)$). This relation is familiar
from \cite{Tak-Tak,gravity}. It is consistent with the
hierarchy and selects a solution that, as $n>1$, is
different from the one selected by (\ref{string1}).
The Lax-Sato equations (\ref{HJ9a}) -- (\ref{J4}) for
the Lax function $\lambda$ and the Hirota
equation (\ref{Hirota}) hold true as they stand
provided the times are redefined as in (\ref{111}).
The solution to the Laplacian growth problem
in the channel geometry (in the
Hele-Shaw cell) can be obtained
from the above formulas as a somewhat tricky limit
$n\to \infty$.
It is unclear, however, if the existing finite-dimensional
solutions of the LGE [4-6] will survive after these transformations.
We should also mention that
inclusion of the ``no-flux'' boundary conditions
at the walls of the wedge, which state that
$\partial\log z / \partial \log w$
is real, imposes an extra symmetry in the solution and
doubles the number of
wedges. We will elucidate these questions in the near future.

At last we point out that the above formulas make sense
for negative integer $n$, too. In particular, the case
$n=-1$ describes the {\it internal} radial problem
when oil is inside while water is outside.
In this case the internal and external
harmonic moments are interchanged in their role:
the internal moments (\ref{h2}) together with
the $C_0$ become independent
variables (cf.\,(\ref{111})) while the external
moments are found as derivatives of the prepotential function
according to (\ref{F}).

\vspace{3mm}

These results were reported at the XIII NEEDS Workshop
(Crete, June 1999).
We are very grateful to the organizers
for the invitation and for the opportunity
to share these results in such a fruitful and nice atmosphere.
We are indebted to Paul Wiegmann for collaboration
in \cite{MWZ,WZ} and many stimulating discussions.
The work of A.Z. was partially supported
by RFBR grant 98-01-00344.

\end{document}